\documentstyle[preprint,aps,psfig]{revtex}
\tightenlines

\begin{document}
\draft
\preprint{}
\title{A Hierarchical Model of Slow Constrained Dynamics}
\author{ 
M.\ A.\ Mu{\~n}oz$^{1}$,
A.\ Gabrielli$^{1,2}$,
H.\ Inaoka$^{1,3}$,
and L.\ Pietronero$^{1}$}
\address{
$^{1}$Dipartimento di Fisica, Universit\`{a} di Roma ``La Sapienza,''\\
Piazzale A.~Moro 2, I-00185 Rome, Italy}
\address{
$^{2}$Dipartimento di Fisica, Universit\`{a} di Roma ``Tor Vergata''\\
Via della Ricerca Scientifica 1, I-00133 Rome, Italy}
\address{
$^{3}$Graduate School of Information Sciences, Tohoku University,\\
Sendai 980-77, Japan}

\maketitle
\begin{abstract}

 We introduce a new simple hierarchically  constrained
 model of slow relaxation. The configurational energy has a simple
form as there is no coupling among the spins defining the system;
the associated stationary distribution is an equilibrium, Gibbsian one. 
However, due to the presence of  hierarchical 
constraints in the dynamics the 
system is found to relax to its equilibrium distribution in an 
extremely slow fashion when suddenly cooled from an initial
temperature, $T_0$, to a final one $T_f$. The relaxation curve in that case
 can be fit
by an stretched exponential curve. On the other hand the relaxation function
 is found to 
be exponential when $T_f >T_0$, with characteristic times depending
on both $T_f$ and $T_0$, with characteristic times obeying an Arrhenius law.
 Numerical results as well as some analytical 
studies are presented. In particular we introduce a simple equation that
captures the essence of the slow relaxation.
   
\end{abstract}
\pacs{PACS number(s): }

\narrowtext

\section{Introduction}
It has been recognized that glasses
show many interesting universal properties
 which have not been satisfactorily explained yet \cite{Angell,Sitges}.
 It is well known that glassy states are not ergodic \cite{Palmer2};
for example, the diffusive motion of molecules in a liquid near the
 glass transition temperature becomes much slower
 than the experimental time scale, and
 the state of the system
 is not able to explore the whole phase space within the observation
 time scale.
The relaxation function, $q(t)$, that describes the decay of the
system energy (or, analogously, of any other relevant magnitude)
towards its stationary state, is observed to evolve
in an extremely slow fashion in the vicinity of
 the glass transition, and
is empirically  found to  be well fit by the Kohlrausch-Williams-Watts
 or stretched-exponential law, i.e.
\begin{equation}
q(t) =  A \exp[-b * (t/\tau)^{\beta}]
\label{stretched}
\end{equation}
with exponent $0 < \beta < 1$ \cite{Bohmer}.
It is often observed that the exponent $\beta$
 decreases with decreasing temperature \cite{Dixon}.

 Nearby the glassy transition standard methods of
statistical mechanics based on equilibrium distributions
 are no longer suitable to describe this kind of systems.
 A   {\it  dynamical approach} is required
for an adequate understanding of slow relaxation processes.

 To the date, several dynamical models have been proposed to get some
insight into 
the nature of the slow 
relaxation  processes in glassy dynamics. 
All of them have in common the presence of some kind of dynamical
frustration, and can be classified in two groups:
those involving
frustration due to the presence of 
 energy and/or entropy barriers (for example spin-glass
type of models \cite{Franz,Ritort1,Bouchaud}),
 and others in which even with a simple
free energy landscape the frustration is directly introduced in the dynamics
\cite{Ritort2}.
 The model we present in this paper belongs to the second type, i.e.
the energy function is very simple, but the dynamics
is strongly constrained; only a reduced number of degrees  of freedom
can evolve freely at a given time step, while the rest remain frozen.

 The new ingredient we include with respect to previously 
studied models is the presence
of a 
{\it hierarchy} of constraints \cite{Evans}.
 With that hierarchy we pretend to mimic
the fact that in slow relaxation processes, usually there are some degrees
of freedom that evolve faster than others and that influence the
dynamics of the slower modes in such a way that a whole hierarchy
 of constrains is generated.

The idea that a theoretical model of slow relaxation
 should be a dynamical and hierarchically constrained one is not
 new. It  was  first
introduced by
 Palmer et.\ al.\ \cite{Palmer}.
They defined a family of models
 consisting of discrete levels of degrees of
 freedom, $n=0,1,2,\ldots$, each level containing $N_{n}$ {\it spins}.
A spin at level $n+1$ can change freely its state
only when $\mu_{n} \leq N_{n}$ spins in level $n$
happen to be at a given specific configuration
 among all the $2^{\mu_{n}}$
 possible ones.
As a consequence of such type of dynamical constraints, 
they argued that
 a stretched-exponential behavior,  Eq.(\ref{stretched}),
 can be reproduced under broad conditions 
\cite{Palmer}. One important point, however, is that in the 
 previous
paper analysis it is implicitly assumed that all the spins above
a given one in the chain influence the dynamics of it, and therefore
the constraints are long-ranged.

 In this paper we consider a very simple microscopic hierarchically 
constrained model for slow relaxation,
 which is a  particular physical realization
of the general scenario introduced in \cite{Palmer}, with two important
differences:

1- Our model does not consider long range dynamical constraints, i.e,
a degree of freedom is constrained directly
 only by the nearest degrees
of freedom in the hierarchy.  As we will show this is enough to generate
a slow relaxation.

2- 
 The main advantage of
this new model is 
that the dynamics and the constraints are specifically defined,
rendering the model suitable to be analyzed using computer 
simulations and detailed mathematical analysis.

 The paper is structured as follows:
 in section 2 we introduce the model, in section 3 we present 
our main numerical results as well as a simple theoretical approximation
that reproduces stretched-exponential type of relaxation; we also 
discuss the nature of the boundary conditions. Finally we 
present the conclusions. A mathematical formulation of the model
and a mean-field like solution are presented in an appendix.

\section{The model}

\begin{figure}
\centerline{\psfig{figure=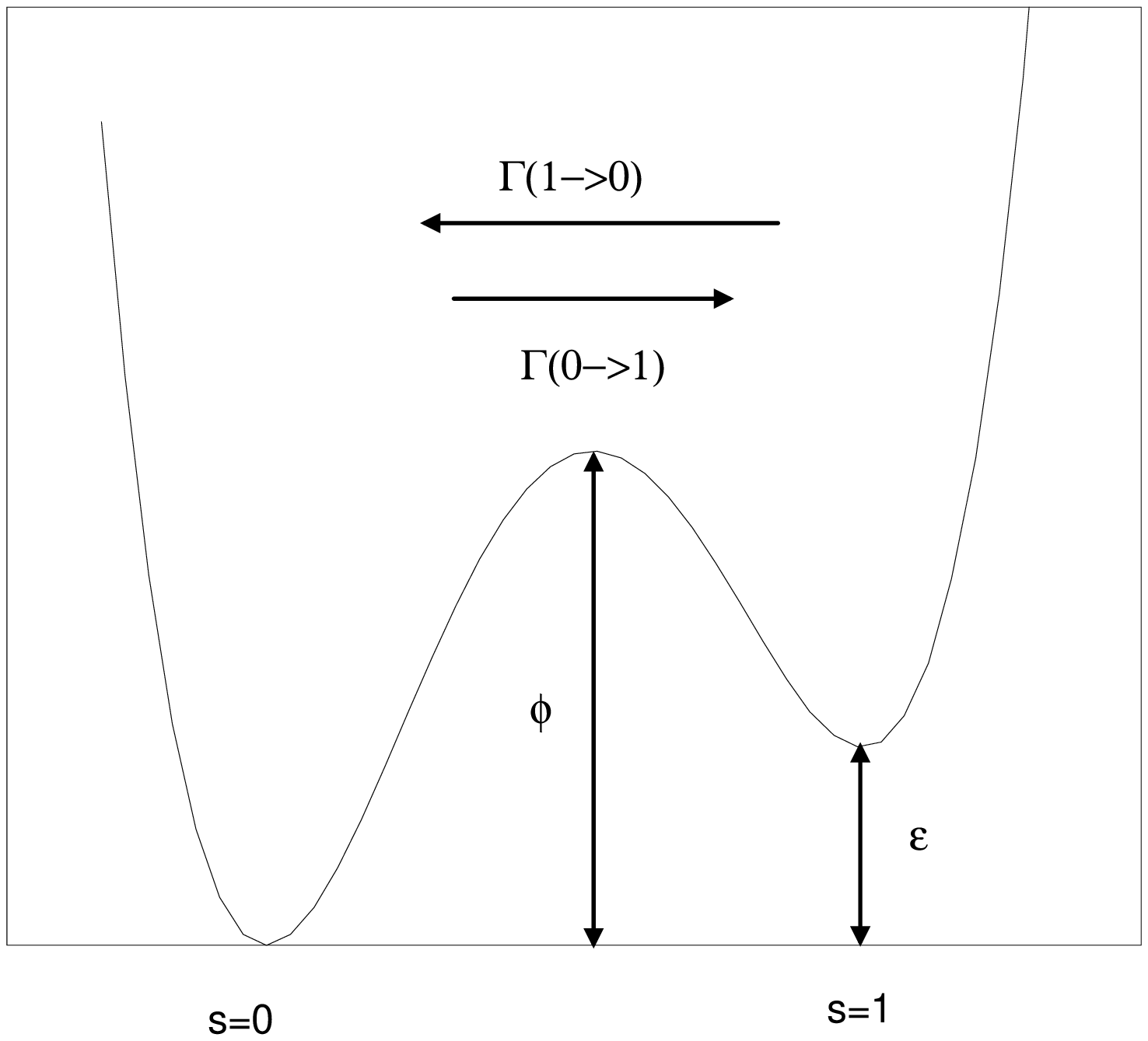,width=10cm}}
\caption{A schematic figure of the
 potential form in which atoms are placed in our model.
Two local minima $s=0$ and 1 are separated by a maximum of height $\phi$.}
\label{potential}
\end{figure}

The model consists of a set of degrees of freedom (or spins),
evolving according to a constrained dynamics. 
The degrees of freedom could,
 in principle, be arranged in different ways, but in 
what follows we consider only a one-dimensional chain with a degree 
at every site.
Each degree of freedom is exposed to the action of an external
 potential which functional
 form is schematically shown in Fig.\ref{potential}.
It has two local minima at positions $s=0$ and
 $s=1$ with values $0$ and $\epsilon$,
 respectively, and a maximum of height $\phi > \epsilon$
 separating the two minima.
We assume that each degree of freedom can be located
 only at the minima of the potential, 
and its state can therefore be characterized
 by a spin-like variable, $s=0,1$.

The spins can change stochastically their state 
with the following 
(Arrhenius) transition rates,
 $\Gamma$
\begin{equation}
\Gamma_{0 \rightarrow 1} = {\tau_{0}}^{-1}
 \exp ( - \frac{\phi}{T_f}) ~~~ , ~~~ 
\Gamma_{1 \rightarrow 0} = 
 {\tau_{0}}^{-1} \exp ( - \frac{\phi - \epsilon}{T_f}) ~~~ ,
\label{gamma}
\end{equation}
where $\tau_{0}$ is a constant that defines a microscopic
 time scale,
 and $T_f$ denotes the temperature.
Assuming that there is no static interaction among spins,
 we can easily calculate the equilibrium probability
 $p_{eq}$ to find a spin at state $s=1$ by using the detailed 
balance condition:
\begin{equation}
p_{eq} = \frac{\Gamma_{0 \rightarrow 1}}
{\Gamma_{0 \rightarrow 1} + \Gamma_{1 \rightarrow 0}}=
{1 \over \exp(\epsilon/T_f) +1}
\label{peq}
\end{equation}
 which gives the Gibbs equilibrium distribution associated to the
potential in Fig. 1 (observe that it does not depend on $\phi$).
In the absence of interaction among spins 
the system exhibits exponential (Debye)
 relaxation towards the equilibrium distribution with characteristic
relaxation time given by:  $\tau=p_{peq} \tau_0 \exp(\phi/T_f)$.

We now introduce a dynamical constraint in the model:
in Fig.\ref{twoatom} (a)  
we show schematically the way in which a
spin, $S_1$, constrains the motion of its right neighbor
$S_2$  \cite{Pietronero}.
When the spin $S_1$ is at the state
  $s_{1}=0$, the dynamics of spin $S_2$
 is obstructed, i.e., $\phi \rightarrow \infty$,
  and jumps
 of $S_2$ between the two states are completely prohibited.
In the case in which both spins $S_1$ and $S_2$
 are initially at state $s=0$,
 for $S_2$ to jump to the state 
 $s_{2}=1$, $S_1$ has to evolve first to
$s_{1}=1$ to clear the path of $S_2$.
Generalizing this rule
the motion 
 of the $(i+1)$-th spin in the chain
 is blocked by the $i$-th spin when
this  one is at state $s_{i}=0$.
Note that while the motion of $s_i$ is affected by $s_{i-1}$
it does not depend on $s_{i+1}$, originating in this way 
a  hierarchy of { \it directed} constraints.
 It is important to observe that the dynamics
is asymmetric, and that even though there is no static
interaction among spins they are {\it dynamically constrained};
therefore the equilibrium distribution is the product of the
single-site equilibrium distributions eq.(\ref{peq}).
This is rigorously proved in the appendix where a 
probabilistic formulation of the model is presented.

Due to this series of constraints
 we expect the system to show slow relaxation.
  One way to see that is in terms
 of the topology of the phase space:
as discussed before, in the two-spin
 case the system cannot
 go directly from $\{s_{1},s_{2}\} = \{0,0\}$ 
to $\{s_{1},s_{2}\}= \{0,1\} $, but
 has to take a roundabout path
 $\{0,0\} \rightarrow \{1,0\}
 \rightarrow \{1,1\} \rightarrow \{0,1\}$.
In the case of $L$-spin system
 the phase space forms a $L$-dimensional hypercube with sides 
of size $1$.
A vertex of the cube represents a
 system state.
Though the maximum geometrical distance
 between any two states is $L^{\frac{1}{2}}$
 measured by the edge length of the hypercube,
 a big fraction of the possible paths
 are obstructed and unavailable.
Therefore the system has
 to pass through more complex roundabout
 paths as the system size $L$ increases, and
we expect those complex trajectories
 to give raise to
 slow relaxation.

 \section{Results}                                  

We have performed numerical simulations
of the model for different system sizes, $L$,
up to $L=1000$, with
open boundary conditions  
\cite{abs}, i.e. the first spin in the chain (hierarchy)
evolves in an unconstrained way (this mimics the fact that
 typically in glasses there are fast, unconstrained, degrees of freedom).
Most of the plots we present are obtained for $L=250$, but the results
have been observed to be very robust when increasing system size.
Simulations have been carried out for both simultaneous and
sequential type of update, and essentially no physical
 difference is observed
among them.
Some parameter values are kept fixed in all the simulations:
 $\tau_{0} = 10.0$, $\phi = 1.0$, and $\epsilon =0.5$.
We have verified that the qualitative general features exhibited by
the model do
not depend on the choice of these values.
As initial condition
 the spins are placed at positions
 $s=0$ or 1 with probabilities corresponding 
to an equilibrium distribution, $p_{eq}(T_0)$  eq.(\ref{peq}), for a given 
initial temperature, $T_0$.  Therefore the dynamics  drives
the system from an equilibrium distribution at $T_0$ to a
different equilibrium distribution at $T_f$.

We introduce the mean energy per spin defined as
\begin{equation}
U(t) \equiv L^{-1} \epsilon \sum_{i=1}^{L} s_{i}(t) ~~~ 
\label{energy}
\end{equation}
(only spins in the $s=1$ state give a non-vanishing contribution 
to the energy).
Given that there is no configurational interaction among spins, 
after sufficiently long times, $U(t)$ approaches its
 equilibrium value $U_{eq} = p_{eq} \epsilon$.
We define a relaxation function $q(t)$ as
\begin{equation}
q(t) \equiv | U(t)-U_{eq} | ~~~ .
\label{defineq}
\end{equation}
which, after a
sufficiently long time, $t$, $q(t)$ approaches zero.

\subsection {General features of the relaxation curves}  

Two typical relaxation curves are shown in figure \ref{Evol};
they correspond to a case in which the system is cooled down,
$T_f < T_0$ (upper one), and in the other the system is heated 
up, i.e., $T_f > T_0$. Their behaviors are essentially different. 

The uppermost corresponds to $1/T_0 = 2.19 $, $1/T_f = 3$ and $L=250$.
 Note that the relaxation is very slow; as the time
 is measured in units of $\tau_0=10$, the maximum time corresponds
to $15000$ time  Monte Carlo steps (the curve is the average of $10^6$
independent runs).
Observe that asymptotically, i.e., after a transient of about
 $t_c \approx 800 \tau_0$
time steps the curve is locally well fit by 
an exponential with a very large characteristic time
 $\tau \sim 2100 \tau_0$, that implies an 
extremely slow relaxation.
 However, it is not clear from numerics  whether
for larger times this exponential behavior will persist, or
the curve will decay even in a slower fashion.
On
the other hand, the whole curve, including the initial faster decay
can be perfectly fit by an stretched exponential with $\beta=0.38$
(in fact the fit in undistinguishable from the numerical data
in Fig. \ref{Evol}). 

  Increasing further the time in the computer simulation 
to decide whether the real asymptotic behavior is an
exponential or an stretched exponential 
is beyond of our available computational power. In any case, for
situations in which the system is cooled down, we always get curves 
that bend progressively in a semilogarithmic plot; and even though
the final part can always be fit with a straight line (exponential
behavior) in no case it is evident whether that exponential
fit gives the right asymptotic behavior. Even for extremely long times
we have this type of ambiguity. However, as for any "reasonable" 
time a stretched exponential can always be fit, we admit 
the relaxation to be non-exponential is this case.  At any rate,
the relaxation is extremely slow in the cooling case.

   The lower curve in figure \ref{Evol} corresponds
to the same initial temperature, $1/T_0 = 2.19 $, and system 
size, than the upper one, but a 
larger final temperature: $1/T_f = 1$.  In this case the transient
is much smaller, $t_c  \approx 60 $, and after it, a pretty clear
exponential behavior settles in. 

  The situation described for the two previous examples is general
for all the relaxation curves: those in which $T_f < T_0$ are well
fit by a stretched exponential, while for the opposite 
situation, $T_f > T_0$, the decay is exponential after a transient.

  We now analyze the transition between the two previous 
regimes in a more quantitative way. In figure \ref{trans}
we show the transient time (i.e. the time after which an 
exponential fit is adequate) as a function of the final
temperature for a fixed initial temperature $1/T_0=2.19$.
 
Note that for very large $T_f$ (small $1/T_f$)
 the system can relax very
fast, there is no effective frustration and $t_c$ is 
small. In fact, for $1/T_f \rightarrow 0$ the
system has a huge degree of thermal activation, and decays
exponentially fast to the equilibrium state with no transient. 
At $T_f=T_0$ the system is already at equilibrium
therefore $t_c=0$.
 Between these two limiting case $t_c$ is larger
than $0$ and behaves in the form shown in Fig. \ref{trans}; observe 
that in this interval $t_c$ is always relatively small, therefore
the asymptotic exponential settles in a short time (smaller than
the observation time).
On the other hand, when $T_f<T_0$ ($1/T_f > 1/ T_0$),
 the behavior is quite different and
$t_c$ grows fastly without bound.
We point out that the previous set of values of $t_c$ for 
$T_f<T_0$ are obtained fixing some maximum time (t=1500). If that time is 
diminished the apparent $t_c$ decreases, while it seems to increase
for longer times. In this way the values plotted in  Fig. \ref{trans} 
are lower bounds for the transient times. In fact, as discussed 
above, if $t_c$ converged to a fixed value for $t \rightarrow
\infty$, that would imply an  exponential asymptotic decay;
while a continuously growing $t_c$ would imply a 
stretched  exponential type of  asymptotic decay. Deciding
which of those two possibilities is the right one from numerics is a 
very difficult task given the extreme slowness of the 
relaxation process.

  Let us now ignore the transients and
present a description of the system behavior in terms of
its asymptotic decay for the $T_f > T_0$ case.

For a given pair of fixed initial and
final temperatures, the relaxation function is well fit
(after the transient) by an exponential (as discussed previously)
 with a characteristic
time $\tau$ that {\it depends on both the initial and
the final temperature}.   

 In Fig. \ref{arr1} and \ref{arr2} we present the results of
our simulations for a chain of length $L=250$ (the results do not
change significatively with increasing system size).

  In Fig. \ref{arr1} the dependence of the characteristic time
 on the final temperature is shown.
The curve
is well fit by an Arrhenius law:  $\tau = C_1 \exp (- C_2/T_f)$
where $C_1$ and $C_2$ are constants.

 In Fig \ref{arr2} we keep fixed the final temperature, $1/T_f=2$
to study the dependence of the characteristic time on the initial
state, we also observe that an Arrhenius law: 
 $\tau = D_1 \exp (- D_2/T_0)$, with $D_1$ and $D_2$
constants holds.
  Therefore the characteristic time depends on both the
final temperature and the initial state.

   We want to point out that in the two previous graphs, 
if we crossed to the other regime (i.e. $T_f < T_0$), plotting
the characteristic time associated to a 
 long time simulation (for example, the one used in figure
 \ref{trans}) we  would  obtain a discontinuous jump:
the characteristic time for a slight system cooling
($T_f = T_0 -\alpha$, with $\alpha$ small and positive)
 is much 
larger (if any) than the one for a slight heating up
($T_f = T_0 +\alpha$).

   Physically the main difference among
 the cooling and the heating processes
is the following:
as can be easily
derived from the equilibrium distribution, the average length, 
$<l>_0$, of chains of spins in the
blocking position, $s=0$ is:
$<l>_0= 1 +\exp(\epsilon/T)$. Observe that it goes to $2$ at
infinite temperature and diverges for vanishing temperature.
Analogously the mean length of $s=1$ chains is
$<l>_1= 1 +\exp(-\epsilon/T)$.
In order to cool the system down the typical length of a chain
of blocking states has to be enlarged. But, of course, the 
dynamics is restricted to sites preceded by islands of $s=1$ states.
 As the number and 
typical length  of these islands with $s=1$
 is decreased in the cooling process
the dynamics becomes slower an slower. 
On the other hand, when the system is heated up, there is no such an
effect: the dynamics is accelerated as more and more
active sites appear in the system configuration. 
In order to render more explicit the previous argument let us perform
 a 
mean field type of calculation:
as we have seen the total energy of the 
system is proportional 
(through $\epsilon$) to the number of spins $N_1(t)$ in the state $1$.
For $N_1(t)$ we can write a mean-field master equation:
\begin{equation}
N_1(t+1)=N_1(t)+(L - N_1(t))\eta_1(t)-N_1(t)\eta_2(t)
\label{N}
\end{equation}
where $L$ is the system size, $L-N_1(t)$ is the number of spins in 
the state $0$ at time $t$,
$\eta_1(t)$ is the probability that at time $t$ a spin in the 
state $0$ flips to $1$ and $\eta_2(t)$ is the probability
of flipping from $1$ to $0$.
If we now divide eq.\ref{N} for $L$, we obtain:
\begin{equation}
 \sigma (t+1) =  \sigma(t) + (1 - \sigma (t)) \eta_1 (t)
  - \sigma(t) \eta_2(t)  
\label{sss}
\end{equation}
where
$\sigma(t)$ is the averaged density of spins
at the $s=1$ state at time $t$.
In this way, eq. (\ref{sss})
 is nothing but an effective balance equation, which establishes
that the probability of being in $s=1$ at time $t+1$ is
given by the probability of being at $s=1$ at time $t$, plus
the probability of being at $s=0$ times the probability
of jumping from $s=0$ to $s=1$, minus the probability
to escape from $s=1$.

We could consider $\eta_1$ and $\eta_2$ as given by some fixed 
values, but in order to keep more track of the original constrained
nature of the dynamics
dynamics we choose them in a slightly different way: 
as $\sigma(t)$ can be interpreted as the probability 
that a fixed spin is in the state $s=1$  at time $t$, we take
$\eta_1=0$ with probability $1- \sigma(t)$, that is the transition
probability is $0$ if the preceding spin is in a blocking state,
 and 
$\eta_1= \exp(-1/T_f)$ (we have taken $\phi=\tau_0=1$) 
with complementary probability $\sigma(t)$, according to 
 eq. (\ref{gamma}).
Analogously,
$\eta_2=0$  with probability $1- \sigma(t)$, $\eta_2=
\exp((\epsilon -1)/T_f)$ with probability $\sigma(t)$.

 In this way, as the system relaxes to its stationary state, the
transition probabilities change and modify the rate at which 
the dynamical variable $\sigma(t)$ evolves.

  The previous simple (but not trivial)
 equation can not be solved analytically,
but it is simple to iterate numerically.
 We have considered
$\sigma(t=0)$ given by its equilibrium value at some temperature
$T_0$, and iterate the equation for different values of the 
temperature $T_f$.  
The result of such
an analysis are shown in figure \ref{SUPER}.  For any set of
parameter values the 
resulting relaxation curve can be fit by an stretched exponential
with exponent $\beta<1$. We will further study the curious analitycal 
properties of this simple approach in a future work.

   Therefore, even in this simple approach,
 in which the spatial structure is almost
completely disregarded, we reproduce a stretched-exponential
type of decay, due to the fact that the jumping probability 
 decays in time as the number
of blocking states grows.

\subsection {Open versus periodic boundary conditions}
 
 A possible way to synthesize a stretched-exponential is by the
convolution of a number of exponential curves with different
characteristic times, $\tau_n$, i.e.
\begin{equation}
q(t) = \sum_{n} w_{n} \exp(-t/\tau_{n})
\label{syn}
\end{equation}
where $w_n$ are some weight factor.    
   
  In order to shed some light on the microscopic origin of that 
slow relaxation process, and clarify 
whether the stretched-exponential behavior describing the transient 
behavior of the relaxation function
comes from a convolution like that of Eq.(\ref{syn}) (with the $n$
in the sum being the position in the chain),
 we have studied
the time relaxation of the energy of the spin
at every site as a function
of its position in the chain.
 In Fig.\ref{decay} we show the
time evolution of $ U(i,t)- U_{eq}$, with $U(i,t)= \epsilon
\langle s_i (t) \rangle$, where
 $\langle \ldots \rangle$ stands for averages over different runs.
It is observed that the first spin in the chain relaxes faster
than the second one, the second one faster than the third, and so on.
I fact, the energies of the first spins $U(i,t)$ relax
 exponentially fast to its equilibrium value,
$U_{eq}= p_{eq} \epsilon$,  with
a time constant, $\tau_i$, that
increases with the position $i$ in the chain. 
 After a certain number of spins that depends on the parameter values,
 the relaxation of every single spin
is indistinguishable from the relaxation observed performing a
simulation with periodic 
boundary conditions and the same parameter values:
 this is what we call the {\it bulk behavior}.
 Therefore, the only difference
between periodic and open boundary conditions is a small effect that 
does not affect the bulk properties. As apparent stretched-exponential 
behavior are observed for the relaxation of the bulk spins when $T_f< T_0$,
 we conclude that
it is not due to a convolution of exponential functions with different
characteristic times associated to the different positions in the chain.

  An alternative possible way to understand the apparent stretched
exponential behavior as associated to a convolution of exponentials
with different characteristic times is by assuming that the islands of
spins in position $s=1$ in the system relax in a different way depending
on how many blocking spins are placed in the immediately superior 
positions in the chain. In this way it is clear that spins blocked 
by only one spin will relax much faster than spins preceded by long chains
of blocking spins. This mechanism would give rise to different relaxation
times for spins in different relative positions in the chain, and therefore
to a global stretched-exponential behavior.

\section{Conclusions}
 
 We have presented a simple model of slow relaxation. Its stationary 
equilibrium distribution is a simple Gibbs distribution, but its
dynamics is strongly constrained. That gives rise to slow relaxation
processes in the case in which the system is cooled from an
initial temperature, $T_0$ to a smaller $T_f$.
The asymptotic behavior of the relaxation
function, after a transient,  is found to be exponential
when heating the system up, with a 
relaxation time depends on both the initial and the final
temperatures in a non-trivial way.
When cooling the system down it is not clear whether the relaxation 
function reaches an exponential behavior asymptotically or if it is 
a stretched-exponential even asymptotically.
In any case, for small enough final temperature, and for large
initial temperatures, the system is found to show a extremely
slow relaxation originated by the constrained dynamical rules.
In this way we put forward, using a very simple model, how 
constrained dynamics can slow down relaxation processes 
in quite a dramatic way. We have also introduced a very simple
 one-variable equation that captures the essence of the slow
relaxation and reproduces stretched-exponential type of decay.

\section{Acknoledgements}

We thanks C. A. Angell, U.\ Marini Bettollo,
 Sebastiano Carpi, and G. Parisi for useful
discussions. We are grateful to the
anonymous referee whose comments and criticisms helped us
to improve notoriously this paper.
This work was partially supported by the
Japan Society for the Promotion of Science
through a grant to H.I.\ and by the
European Union through a grant to M.A.M.

\section{Appendix: Analytical approach}

  Now we present an analytical attempt to understand the previously
described properties.
 Our model can be formally represented in terms of a Markov chain,      
defined by the following equation:
\begin{equation}
P(\{s\};t) = \sum_{\{{s'}\} } W( \{{s'}\} \rightarrow
\{s\} ) P(\{{s'}\};t-1)
\label{eq1}
\end{equation}
where $ \{s\} = \{ s_i = s_i(t) \}$,
 and  $ \{s'\} = \{ {s'}_i = s_i(t-1) \}$ for every $i=1,2,..., L$, are
system configurations at time $t$ and $t-1$ respectively. The  equation
\ref{eq1}
 states how the probability, $P(\{s\};t)$
 of finding the system at a given time $t$ in a configuration $\{s\}$ 
evolves in time. The transition probabilities, $W( \{{s'}\} \rightarrow
\{s\} ) P(\{{s'}\};t-1)$ are given by:
\begin{equation}
W( \{{s'}\} \rightarrow \{s\} )
 =  \prod_{i=1}^{L} \omega( {s'}_i  \rightarrow s_i ; {s'}_{i-1}  )
\label{eq2}
\end{equation}
with
\begin{equation}
\omega( {s'}_i  \rightarrow s_i ; {s'}_{i-1}  )
=
\delta_{ {s'}_i , s_i   } - \delta_{ {s'}_{i-1} , 1   }
(-1)^{ {s'}_i + s_i} {\exp( \beta \epsilon {s'}_i ) \over
\tau_0 \exp( \beta \phi ) }
\label{eq3}
\end{equation}
and the boundary condition $s_0(t)=1$.
 This is nothing but the mathematical expression of
the transition probabilities 
described in the previous section.

By direct substitution it is easy to verify
that the equilibrium distribution 
\begin{equation}
P(\{s\};eq) = \prod_{i=1}^{L}\frac{\exp{ (-\beta \epsilon s_i)}}
{1+\exp{(-\beta \epsilon)}}
\label{equil}
\end{equation}
 is the stationary solution of Eq.(\ref{eq1}).

 From the general equation Eq.(\ref{eq1}),
 we can derive a hierarchy of equations
for the $m$-body probability distributions
 (similar to the BBGKY hierarchy in statistical mechanics), 
In particular, for $m=1$ 
\begin{equation}
p(s_i;t) = \sum_{ {s'}_{i-1}=0}^{1}    \sum_{ {s'}_{i}=0}^{1}
\omega( {s'}_i  \rightarrow s_i ; {s'}_{i-1}  )
p({s'}_{i-1} , {s'}_i ; t-1)
\label{eq4}
\end{equation}
where
\begin{equation}
p(s_i;t) = \sum_{s_1=0}^{1} ... \sum_{s_{i-1}=0}^{1}
\sum_{s_{i+1}=0}^{1} ... \sum_{s_L=0}^{1} P( \{ s\}; t)
\label{eq4b}
\end{equation}
are the one-body probability functions at time $t$, 
\begin{equation}
p({s'}_{i-1} , {s'}_i ; t-1) =
 \sum_{{s'}_1=0}^{1} ... \sum_{{s'}_{i-2}=0}^{1}
\sum_{{s'}_{i+1}=0}^{1} ... \sum_{{s'}_L=0}^{1} P( \{ {s'}\}; t-1)
\label{eq4c}
\end{equation}
are the two-body probability functions at time $t-1$.
  Analyzing this set of equations is a difficult task as can
be seen from the fact that the one-body probability functions
depend on the two-body probabilities distributions
 $ p({s}_{i-1} , {s}_i ; t-1)$; the equations for these depend 
on three-body probabilities and so on. 

  The first analytical approach ( the only one we complete
in this paper) consists of approximating the two-body
probabilities by the product of two one-body probability functions,
namely:
\begin{equation}
 p({s'}_{i-1} , {s'}_i ; t-1) \approx  p({s'}_{i-1} ; t-1)
    p({s'}_i ; t-1) ~~~ .
\label{eq5}
\end{equation}
This is a mean field like approximation given that high order
 correlation  
are neglected.
Introducing Eq.(\ref{eq5}) in Eq.(\ref{eq4}), and after some simple algebra, 
we get
\begin{equation}
p_i(t)-p_i(t-1)= { 1 \over \tau_0 \exp( \beta \phi)} [
p_{i-1}(t-1)( 1- p_i(t-1)) -   \\
\exp( \beta \epsilon)
p_{i-1}(t-1) p_i(t-1) ]
\label{eq6}
\end{equation}
where $p_i(t)= p( s_i=1;t)$.
In the case of an open chain we put $p_0(t)=1$, which 
corresponds to the
fact that the first spin is unconstrained at any time.

   In the continuous time limit, i.e.\ when $\tau_0 \gg 1 $ (which 
is the case in the numerical simulations), the previous equation 
can be written as:

\begin{equation}
{ d p_i(t) \over dt } = { 1 \over \tau_0 \exp( \beta \phi)}[
p_{i-1}(t)( 1- p_i(t)) - \exp( \beta \epsilon)
p_{i-1}(t) p_i(t) ]
\label{eq6b}
\end{equation}
which general solution is
\begin{equation}
p_{i}(t) = p_{eq} +(p_{i}(0)-p_{eq})\exp(I_{i}(t)/\tau_{1}) ~~~ ,
\label{solution}
\end{equation}
with $\tau_{1} = \tau_{0} p_{eq} \exp(\phi/T_f)$
(which is the free relaxation time) and 
$I_{i} = \int_{0}^{t} p_{i-1}(t^{\prime})
 \, dt^{\prime}$ is an effective time for $i$-th spin.
Substituting $I_{1} = t$ in Eq.(\ref{solution}), 
 we can calculate $I_{2}$. It is obtained
 to be given asymptotically by
 $I_{2} \simeq p_{eq} t$ for $t \gg \tau/p_{eq}$, and therefore
 Eq.(\ref{solution}), 
 for $i=2$ can be rewritten as
\begin{equation}
p_{i}(t) = p_{eq} +(p_{i}(0)-p_{eq})\exp(-p_{eq} t / \tau_{1}) ~~~ .
\label{solutionapp}
\end{equation}
By iterating the same procedure we get
$I_{i} \simeq p_{eq}t$
 and Eq.(\ref{solutionapp}) as the solution for $p_{i}$ for any $i>1$.
The relaxation time for $i$-th spin $\tau_{i}$ by
 Eq.(\ref{solutionapp}) is
\begin{equation}
\tau_{i} = \tau_{1} / p_{eq} \sim
 \exp(\mbox{const.} \times \frac{1}{T_f}) ~~~ ,
\label{taubymaster}
\end{equation}
which has a temperature dependence as that of strong liquids. However,
this mean field approximation does not reproduce 
correctly the characteristic decay times.
  We expect that more accurate approximations
to Eq.(\ref{eq1}) reproduce 
more accurately the behavior of the relaxation function. Results 
obtained by truncating the hierarchy at higher levels will be presented
elsewhere \cite{noi}.

\begin{figure}
\caption{   \label{twoatom} When the spin 1 is placed at position
 $s_{1}=0$ the transition of the spin 2 is prohibited.}
\end{figure}

\begin{figure}
\caption{ \label{twoatom2} The phase space for the two spin system.
The direct path from $\{s_{1},s_{2}\} =
 \{0,0\}$ to $\{0,1\}$ is obstructed
 and the transition between these two states is prohibited.}
\end{figure}

\newpage

\begin{figure}
\centerline{\psfig{figure=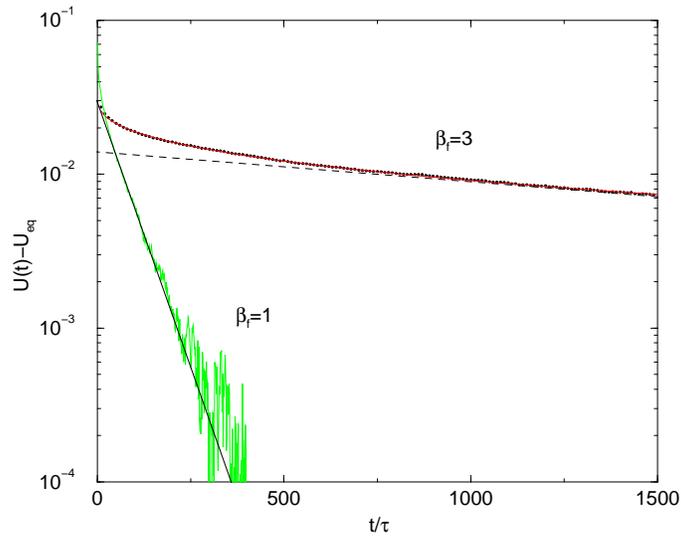,width=10cm}}
\caption
{\label{Evol}  Typical relaxation curves for $T_f <T_0$ (uppermost,
$1/T_f = 3$ and $1/T_0 = 2.19 $), and $T_f > T_0$ (lowermost 
$1/T_f = 1$ and $1/T_0 = 2.19 $). The second one is exponential while
the first one can be fit by an stretched exponential with $\beta=0.38$;
an exponential fit (dashed line) is valid for times larger than 
$\approx 800$.
$L=250$.
}
\end{figure}

\begin{figure}
\centerline{\psfig{figure=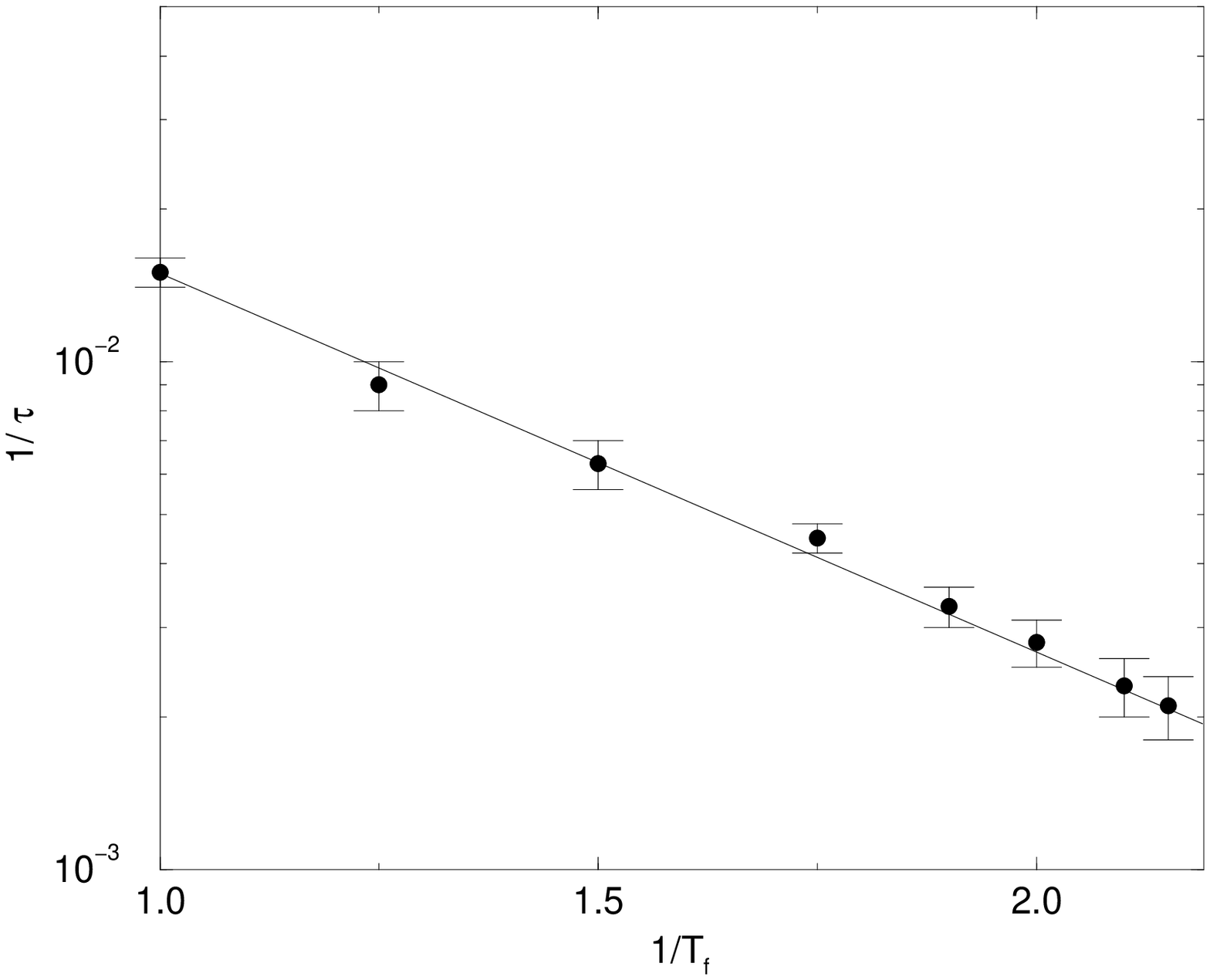,width=10cm}}
\caption
{\label{arr1}
Semilog plot of the inverse of the characteristic time as a function
of $1/T_f$ for a fixed $1/T_0= 2.19$. $L=250$.
}
\end{figure}
 
\newpage       

\begin{figure}
\centerline{\psfig{figure=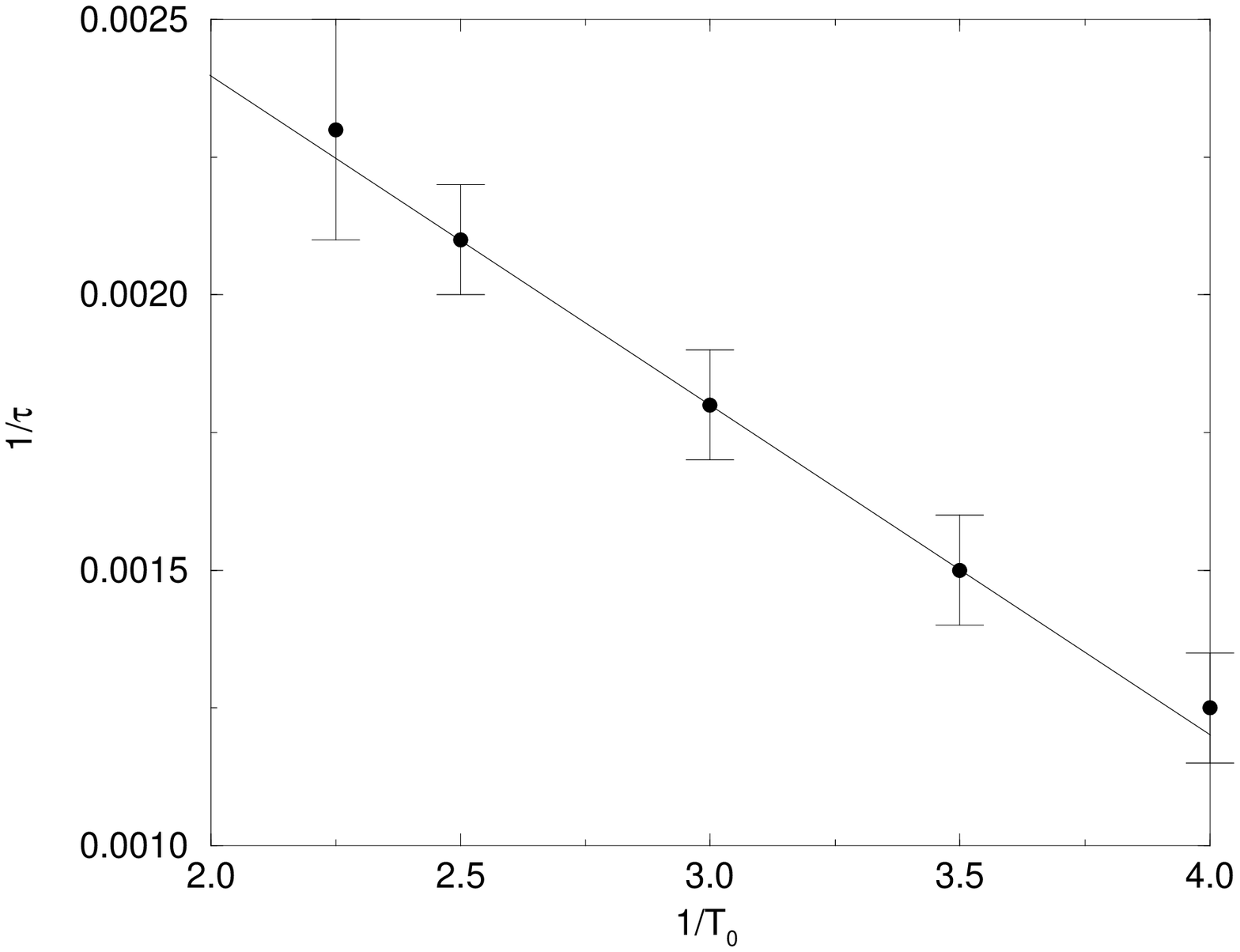,width=10cm}}
\caption
{ \label{arr2}
Semilog plot of the inverse of the characteristic time as a function
of $1/T_0$ for a fixed $1/T_f=2$. $L=250$.
}
\end{figure}

\begin{figure}
\centerline{\psfig{figure=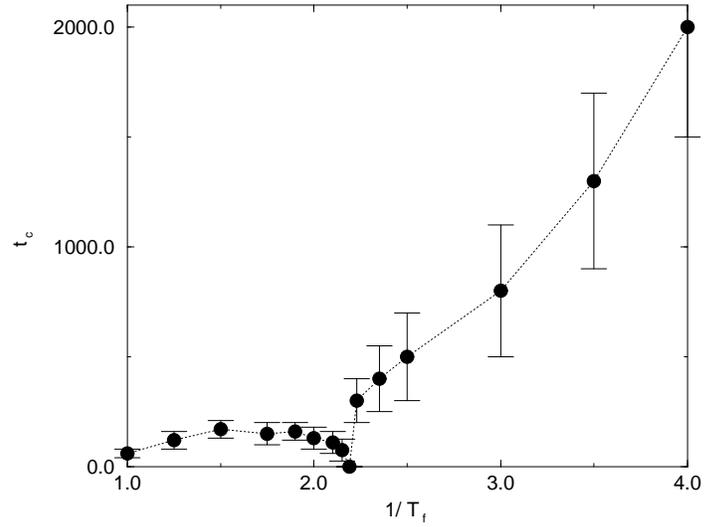,width=10cm}}
\caption
{ \label{trans} Transient time (as defined in the text) as a function
of $1/T_f$, with $1/T_0=2.19$. $L=250$.
}
\end{figure}

\newpage

\begin{figure}
\centerline{\psfig{figure=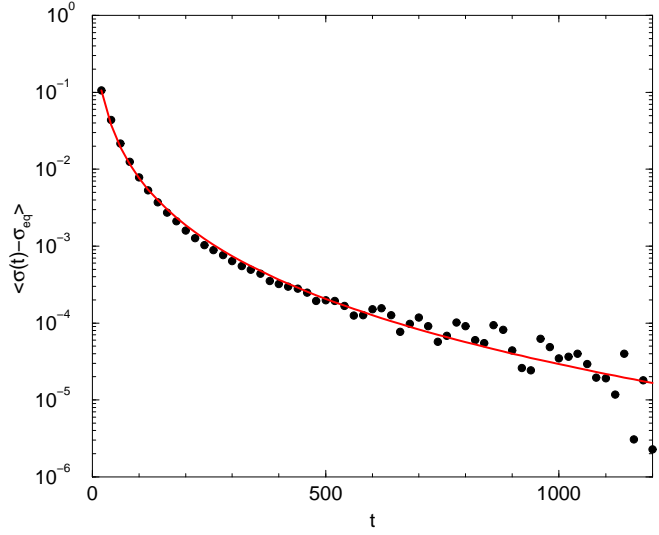,width=10cm}}
\caption
{\label{SUPER} Stretched exponential decay associated to eq.(7),
when the system relaxes  with $1/T_f=3$ from an initial 
$\sigma(0)= 1/(1+exp(1/2))$. 
The fit is given by $b=3.7535$ and $\beta=0.201$ (see eq.(1)).
}
\end{figure}

\begin{figure}
\centerline{\psfig{figure=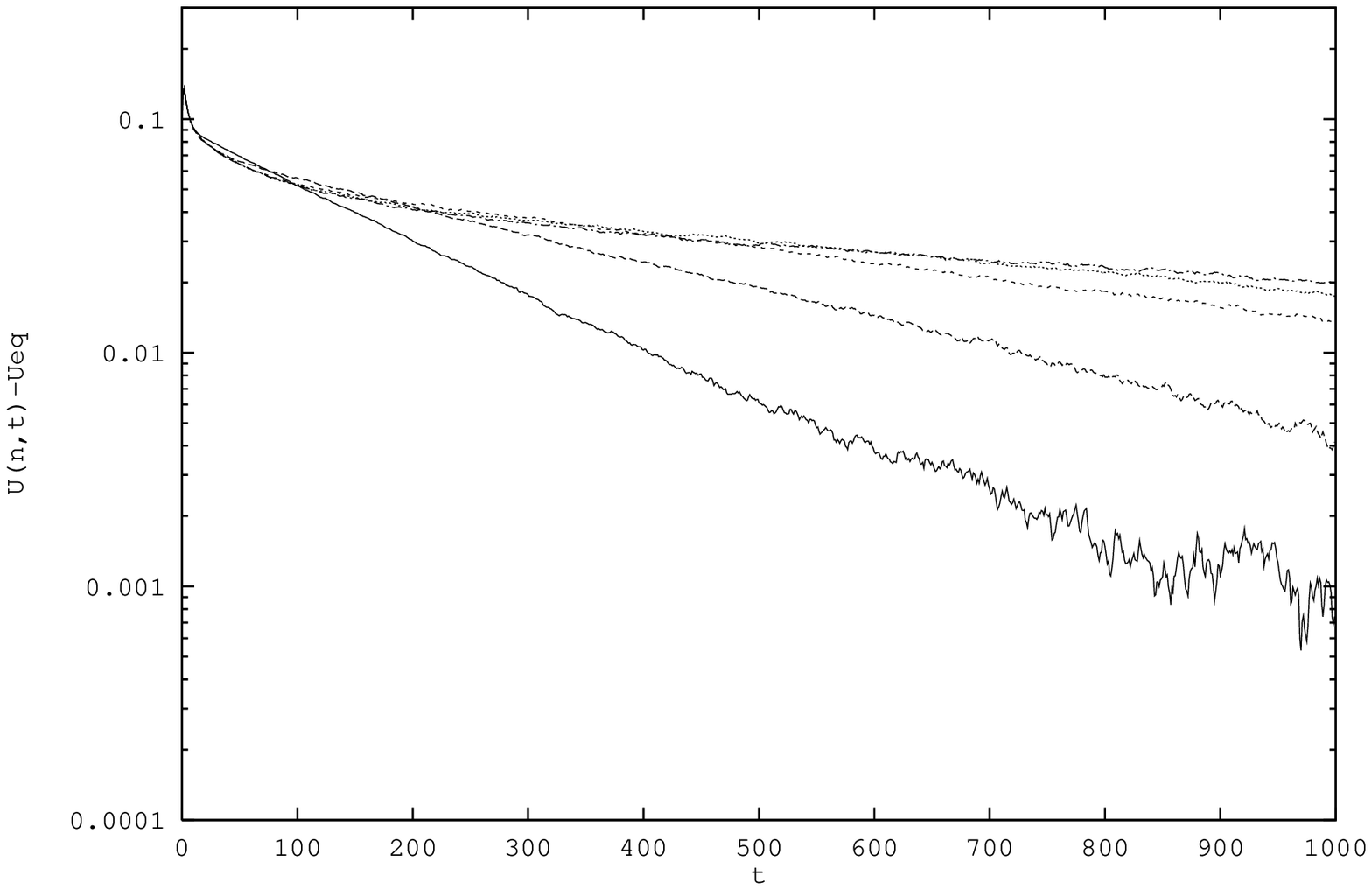,width=10cm}}
\caption{   \label{decay} Semilog plot of the decay of $U(i,t)-U_{eq}$
 for the first seven spins in an open chain
as a function of time, for a system with $L=250$.
 The lowermost curve corresponds to the first spin in
the chain, the next one to the second one and so on. }
\end{figure}


\begin{references}



\bibitem{Angell}
C.\ A.\ Angell, Science {\bf 267}, 1924 (1995) and references there in.

\bibitem{Sitges} {\it Complex behaviour of Glassy Systems}, 
Ed. M. Rub{\'\i} and C. P\'erez-Vicente, Lecture Notes in Physics,
{\bf 492}, 1997.


\bibitem{Palmer2}
R.\ G.\ Palmer, Adv.\ Phys.\ {\bf 31}, 669 (1982).

\bibitem{Bohmer}
R.\ B\"{o}hmer, K.\ L.\ Ngai,
 C.\ A.\ Angell, D.\ J.\ Plazek, J.\ Chem.\ Phys.\ {\bf 99}, 4201 (1993).



\bibitem{Dixon}
P.\ K.\ Dixon and S.\ R.\ Nagel, Phys.\ Rev.\ Lett.\ {\bf 61}, 341 (1988).
                

\bibitem{Franz} S. Franz and G. Parisi, J. Phys. I France,
{\bf 5}, 1401 (1995).

                                                       
\bibitem{Ritort1}  F. Ritort, Phys. Rev. Lett. {\bf 75}, 1190 (1995).
 

\bibitem{Bouchaud} J. P. Bouchaud and M. Mezard, J. Phys. I France,
{\bf 4}, 1109 (1994)

\bibitem{Evans} G. W Evans, Rev. Mod. Phys. {\bf 65}, 1281 (1993). 

\bibitem{Ritort2} E. Follana and F. Ritort, Phys. Rev. {\bf B 54}, 930
(1996).

\bibitem{Evans} Models that present some resemblance with the
one we introduce here have been previously studied in a different context;
see for instance,
G. W Evans, Rev. Mod. Phys. {\bf 65 }, 1281 (1993).
                        
\bibitem{Palmer}
R.\ G.\ Palmer, D.\ L.\ Stein, E.\ Abrahams,
 and P.\ W.\ Anderson, Phys.\ Rev.\ Lett.\ {\bf 53}, 958 (1984).


\bibitem{Pietronero}
L.\ Pietronero, in {\em Fractals in Physics},
 edited by L.\ Pietronero and E.\ Tosatti
 (Elsevier Science Publishers B.\ V., 1986), pp.\ 417.
                                                                  

\bibitem{abs} Note that with periodic boundary conditions
the system could reach an absorbing state, $s_i=0$, for all $i$,
from which the system cannot escape.

                                                             
\bibitem{noi} A. Gabrielli, M.A. Mu{\~n}oz and L. Pietronero, in preparation.






\end{references}
\end{document}